# Motion of half-spin particles in the axially symmetric field of naked singularities of static q-metric

V.P.Neznamov[1,2*], V.E.Shemarulin[1†]

[1]RFNC-VNIIEF, Russia, Sarov, Mira pr., 37, 607188
[2]National Research Nuclear University MEPhI, Moscow, Russia

## Abstract

Quantum-mechanical motion of a half-spin particle was examined in the axially symmetric field of static naked singularities formed by mass distribution with quadrupole moment (q-metric). The analysis was performed by means of the method of effective potentials of the Dirac equation generalized for the case when radial and angular variables are not separated. As $-1 < q < q_{\lim}$, $|q_{\lim}| \ll 1$ the naked singularities do not except the existence of stationary bound states of Dirac particles for a prolate mass distribution in the q-metric along the axial axis.

For the oblate mass distribution, the naked singularities of the q-metric are separated from the Dirac particle by infinitely large repulsive barriers with the subsequent potential well deepening while moving along the angle from the equator (or from $\theta = \theta_{\min}$, $\theta = \pi - \theta_{\min}$) towards poles. The exception are the poles and, as $0 < q < q^*$, some points $\theta_i$ for the states of the particle with $j \geq 3/2$.

---

[*] E-mail: neznamov@vniief.ru
[†] E-mail: shemarulin@vniief.ru

**Introduction**

In terms of multipole moments, the simplest static solution to Einstein vacuum equations is the Schwarzschild metrics with a mass monopole moment only. The first vacuum solution with the quadrupole mass moment was obtained by Weyl in 1917 [1]. Ever since, many papers devoted to study of vacuum solutions with non-zero multipole moments have appeared in literature (see, for instance, [2] - [14]). Rather a simple compact form for the quadrupole metric (q-metric) was obtained in [4]. In spherical coordinates, it can be represented as

$$ds^2 = \left(1-\frac{r_0}{r}\right)^{1+q} c^2 dt^2 - \left(1-\frac{r_0}{r}\right)^{-q}\left[\left(1+\frac{r_0^2 \sin^2\theta}{4r^2\left(1-\frac{r_0}{r}\right)}\right)^{-q(2+q)}\left(\frac{dr^2}{1-\frac{r_0}{r}}+r^2 d\theta^2\right)+r^2\sin^2\theta d\varphi^2\right]. \quad (1)$$

In (1), $r_0 = \frac{2GM}{c^2}$ is the event horizon (the gravitational radius) of the Schwarzschild field. Below, we are going to use the system of units $\hbar = c = 1$.

The q-metric is the axial-symmetric vacuum solution which as $q \to 0$ is reduced to the spherical-symmetric Schwarzschild metric.

From the positivity condition of the Arnowitt-Deser-Misner mass, the condition $q > -1$ follows [13]. The interval $q \in (-1, 0)$ describes prolate mass distribution of the q-metric source along the axial axis; the interval $q \in (0, \infty)$ describes the oblate mass distribution.

The q-metric has naked singularities as $r = 0$ and $r = r_0$. At some parameter values there exists the third singularity [13], determined by the equation

$$r^2 - r_0 r + \frac{r_0^2}{4}\sin^2\theta = 0. \quad (2)$$

Our paper is devoted to the study of quantum-mechanical motion of half-spin particles in the field of naked singularities of the q-metric (1).

The analysis was performed by means of effective potentials of the Dirac equation in the q-metric field. For such an analysis, the self-conjugate Hamiltonian was derived and the method of effective potentials was generalized for the case when radial and angular variables are not separated. As a result, it was shown that as $-1 < q < q_{\lim} < 0$ the naked singularities do not prevent the possibility of existence beyond its stationary bound states of Dirac particles for the prolate mass distribution in the q-metric ($|q|_{\lim} \ll 1$, the value $q_{\lim}$ depends on the parameters of the q-metric (1)).

For the oblate mass distribution, the naked singularities of the q-metric are separated from the Dirac particle by infinitely high repulsive barriers with the subsequent potential well deepening while moving along the angle from the equator (or from $\theta = \theta_{min}$, $\theta = \pi - \theta_{min}$) towards the poles.

The exception are the poles and, as $0 < q < q^*$, some points $\theta_i$ for the states of the particle with $j \geq 3/2$. (The calculations performed by means of the software package "Maple" have shown that $1.4142 < q^* \approx \sqrt{2} < 1.41424$. The detailed description of $q^*$ is given in 3.1).

The paper is organized as follows: In section 1, the self-conjugate Dirac Hamiltonian in the q-metric field is derived. In section 2, for the case when radial and angular variables are not separated the method for obtaining effective potentials of the Dirac equation was generalized. In section 3, the obtained effective potential is examined depending on $(r,\theta)$ and initial parameters of the q-metric. In section 4, the compliance of the obtained results with the hypothesis of cosmic censorship is discussed. In Conclusions, the obtained results are briefly discussed.

## 1. Self-conjugate Hamiltonian of a half-spin particle in the q-metric field

The required Hamiltonian is determined by means of algorithms for obtaining self-conjugate Dirac Hamiltonians in the exterior gravitational fields by using the methods of pseudo-Hermitian quantum mechanics [15] - [17].

In (1), let us denote

$$f_S = 1 - \frac{r_0}{r}, \tag{3}$$

$$a(r,\theta) = \left(1 + \frac{r_0^2 \sin^2\theta}{4r^2 f_S}\right)^{-q(2+q)}. \tag{4}$$

Then, the non-zero components of the metric tensor in (1) are

$$g_{00} = f_S^{1+q}; \quad g_{11} = -f_S^{-1-q} a(r,\theta); \quad g_{22} = -f_S^{-q} a(r,\theta) r^2; \quad g_{33} = -f_S^{-q} r^2 \sin^2\theta. \tag{5}$$

The nonzero tetrad vectors in the Schwinger gauge [18] and the Dirac $\gamma$-matrix with the global indices are

$$\tilde{H}^0_{\underline{0}} = f_S^{-\frac{1+q}{2}}; \quad \tilde{H}^1_{\underline{1}} = \frac{f_S^{\frac{1+q}{2}}}{a(r,\theta)^{1/2}}; \quad \tilde{H}^2_{\underline{2}} = \frac{f_S^{\frac{q}{2}}}{a(r,\theta)^{1/2} r}; \quad \tilde{H}^3_{\underline{3}} = \frac{f_S^{\frac{q}{2}}}{r\sin\theta}, \tag{6}$$

$$\tilde{\gamma}^{\underline{0}} = f_s^{-\frac{1+q}{2}} \gamma^{\underline{0}}; \quad \tilde{\gamma}^{\underline{1}} = \frac{f_s^{\frac{1+q}{2}}}{a(r,\theta)^{1/2}} \gamma^{\underline{1}}; \quad \tilde{\gamma}^{\underline{2}} = \frac{f_s^{\frac{q}{2}}}{ra(r,\theta)^{1/2}} \gamma^{\underline{2}}; \quad \tilde{\gamma}^{\underline{3}} = \frac{f_s^{\frac{q}{2}}}{r\sin\theta} \gamma^{\underline{3}}. \tag{7}$$

In (6), (7), the underlined indices are local indices. The sign «~» over quantities means that they are calculated using tetrads in Schwinger gauge.

For diagonal metric tensors $g_{\mu\nu}$, the self-conjugate Hamiltonian in $\eta$-representation (with a plane scalar product of wave functions) is easily derived from the equality obtaining in [17]

$$H_\eta = \frac{1}{2}\left(\tilde{H}_{red} + \tilde{H}_{red}^+\right), \tag{8}$$

where

$$\tilde{H}_{red} = \frac{m}{g^{00}} \tilde{\gamma}^{\underline{0}} - \frac{i}{g^{00}} \tilde{\gamma}^{\underline{0}} \tilde{\gamma}^{\underline{k}} \frac{\partial}{\partial x^k}. \tag{9}$$

In (8), the sign «+» means the Hermitian conjugation.

In (9), $m$ is a mass of the Dirac particle, $g^{00}$ is a component of inverse metric tensor.

Taking into account (5), (7), we obtain

$$H_\eta = f_s^{\frac{1+q}{2}} m\gamma^{\underline{0}} - if_s^{\frac{1+q}{2}} \frac{1}{a^{1/2}} \gamma^{\underline{0}}\gamma^{\underline{1}}\left(\frac{\partial}{\partial r} + \frac{1}{r}\right) - \frac{i}{2}\gamma^{\underline{0}}\gamma^{\underline{1}} \frac{\partial}{\partial r}\left(\frac{f_s^{1+q}}{a^{1/2}}\right) - if_s^{\frac{1+2q}{2}} \frac{1}{ra^{1/2}} \gamma^{\underline{0}}\gamma^{\underline{2}}\left(\frac{\partial}{\partial\theta} + \frac{1}{2}\mathrm{ctg}\,\theta\right) - \\ -\frac{i}{2} f_s^{\frac{1+2q}{2}} \frac{1}{r}\gamma^{\underline{0}}\gamma^{\underline{2}} \frac{\partial}{\partial\theta}\left(\frac{1}{a^{1/2}}\right) - if_s^{\frac{1+2q}{2}} \frac{1}{r\sin\theta}\gamma^{\underline{0}}\gamma^{\underline{3}} \frac{\partial}{\partial\varphi}. \tag{10}$$

The Dirac equation in the Hamiltonian form for a half-spin particle in the field of the naked singularities of the q-metric has the form of

$$i\frac{\partial \Psi(\mathbf{r},t)}{\partial t} = H_\eta \Psi(\mathbf{r},t). \tag{11}$$

## 2. Effective potentials for the field of the naked singularities of the q-metric

It is seen from the Hamiltonian (10) that the radial and angular variables $(r,\theta)$ in the equation (11) are not separated. The generalization of the standard method is needed to obtain effective potentials by means of squaring Dirac equations for real radial wave functions.

Let us represent the wave function $\Psi(\mathbf{r},t)$ in (11) as

$$\Psi(\mathbf{r},t) = \begin{pmatrix} \varphi(r,\theta)\cdot\xi(\theta) \\ -i\chi(r,\theta)\sigma^3 \cdot \xi(\theta) \end{pmatrix} e^{-iEt} e^{im_\varphi \varphi}. \tag{12}$$

In (12), $E$ is Dirac particle energy, $m_\varphi$ is magnetic quantum number, the spinor

$$\xi(\theta) = \begin{pmatrix} _{-1/2}Y(\theta) \\ _{+1/2}Y(\theta) \end{pmatrix}$$ represents spherical harmonics for a half spin. The explicit form $\xi(\theta)$ can be represented as [19]

$$\xi(\theta) = \begin{pmatrix} _{-1/2}Y_{jm_\varphi}(\theta) \\ _{1/2}Y_{jm_\varphi}(\theta) \end{pmatrix} = (-1)^{m_\varphi + 1/2} \sqrt{\frac{1}{4\pi} \frac{(j-m_\varphi)!}{(j+m_\varphi)!}} \begin{pmatrix} \cos\theta/2 & \sin\theta/2 \\ -\sin\theta/2 & \cos\theta/2 \end{pmatrix} \times$$
$$\times \begin{pmatrix} \left(\kappa - m_\varphi + \frac{1}{2}\right) P_l^{m_\varphi - 1/2}(\theta) \\ P_l^{m_\varphi + 1/2}(\theta) \end{pmatrix}. \tag{13}$$

In (13), $P_l^{m_\varphi \pm 1/2}$ are associated Legendre functions. $j, l$ are quantum numbers of the total angular and orbital momentum of a Dirac particle.

Then, let us note the following:

    1. Since the variables $(r,\theta)$ in (11) are not separated, the functions $\varphi(r,\theta)$ and $\chi(r,\theta)$ depend on $r$ and $\theta$.

    2. In order to obtain real effective potentials, it is necessary for the functions $\varphi(r,\theta)$ and $\chi(r,\theta)$ to be real as well.

Upon substitution (12), the equation (11) will contain spinors $\xi(\theta)$, $\frac{d\xi(\theta)}{d\theta}$, functions $\varphi(r,\theta)$, $\chi(r,\theta)$ and their derivatives with respect to $r$ and $\theta$.

If we make the equivalent substitution in the Hamiltonian (10)
$$\gamma^1 \to \gamma^3, \ \gamma^3 \to \gamma^2, \ \gamma^2 \to \gamma^1, \tag{14}$$

the derivative $\frac{d\xi(\theta)}{d\theta}$ in (11) can be reduced by using the Brill and Wheeler equation [20]

$$\left[i\sigma^2\left(\frac{\partial}{\partial\theta} + \frac{1}{2}\mathrm{ctg}\,\theta\right) + \frac{m_\varphi}{\sin\theta}\sigma^1\right]\xi(\theta) = \kappa\xi(\theta). \tag{15}$$

In (15), $\sigma^1, \sigma^2$ are Pauli matrices;

$$\kappa = \mp 1, \mp 2... = \begin{cases} -(l+1), & j = l + \frac{1}{2} \\ l, & j = l - \frac{1}{2} \end{cases}. \tag{16}$$

As the result, taking into account (12) and definition of the spinor[‡] $\xi(\theta) = \begin{pmatrix} _{-1/2}Y(\theta) \\ _{+1/2}Y(\theta) \end{pmatrix}$, the equation (11) can be written as the system of four equations

$$E\varphi_{-1/2}Y = f_s^{\frac{1+q}{2}} m\varphi_{-1/2}Y - \left[ f_s^{1+q} \frac{1}{a^{1/2}}\left(\frac{\partial}{\partial r} + \frac{1}{r}\right) - f_s^{\frac{1+2q}{2}} \frac{1}{a^{1/2}} \frac{\kappa}{r} + \frac{1}{2}\frac{\partial}{\partial r}\left(f_s^{1+q} \frac{1}{a^{1/2}}\right) \right] \chi_{-1/2}Y +$$
$$+ f_s^{\frac{1+2q}{2}} \frac{1}{a^{1/2}} \frac{1}{r} \frac{\partial}{\partial \theta}(\chi(r,\theta))_{+1/2}Y + \frac{1}{2} f_s^{\frac{1+2q}{2}} \frac{1}{r} \frac{\partial}{\partial \theta}\left(\frac{1}{a^{1/2}}\right) \chi_{+1/2}Y + \frac{m_\varphi}{r\sin\theta} f_s^{\frac{1+2q}{2}} \left(1 - \frac{1}{a^{1/2}}\right) \chi_{+1/2}Y. \quad (17)$$

$$E\varphi_{+1/2}Y = f_s^{\frac{1+q}{2}} m\varphi_{+1/2}Y - \left[ f_s^{1+q} \frac{1}{a^{1/2}}\left(\frac{\partial}{\partial r} + \frac{1}{r}\right) - f_s^{\frac{1+2q}{2}} \frac{1}{a^{1/2}} \frac{\kappa}{r} + \frac{1}{2}\frac{\partial}{\partial r}\left(f_s^{1+q} \frac{1}{a^{1/2}}\right) \right] \chi_{+1/2}Y -$$
$$- f_s^{\frac{1+2q}{2}} \frac{1}{a^{1/2}} \frac{1}{r} \frac{\partial}{\partial \theta}(\chi(r,\theta))_{-1/2}Y - \frac{1}{2} f_s^{\frac{1+2q}{2}} \frac{1}{r} \frac{\partial}{\partial \theta}\left(\frac{1}{a^{1/2}}\right) \chi_{-1/2}Y + \frac{m_\varphi}{r\sin\theta} f_s^{\frac{1+2q}{2}} \left(1 - \frac{1}{a^{1/2}}\right) \chi_{-1/2}Y. \quad (18)$$

$$E\chi_{-1/2}Y = -f_s^{\frac{1+q}{2}} m\chi_{-1/2}Y + \left[ f_s^{1+q} \frac{1}{a^{1/2}}\left(\frac{\partial}{\partial r} + \frac{1}{r}\right) + f_s^{\frac{1+2q}{2}} \frac{1}{a^{1/2}} \frac{\kappa}{r} + \frac{1}{2}\frac{\partial}{\partial r}\left(f_s^{1+q} \frac{1}{a^{1/2}}\right) \right] \varphi_{-1/2}Y +$$
$$+ f_s^{\frac{1+2q}{2}} \frac{1}{a^{1/2}} \frac{1}{r} \frac{\partial}{\partial \theta}(\varphi(r,\theta))_{+1/2}Y + \frac{1}{2} f_s^{\frac{1+2q}{2}} \frac{1}{r} \frac{\partial}{\partial \theta}\left(\frac{1}{a^{1/2}}\right) \varphi_{+1/2}Y + \frac{m_\varphi}{r\sin\theta} f_s^{\frac{1+2q}{2}} \left(1 - \frac{1}{a^{1/2}}\right) \varphi_{+1/2}Y. \quad (19)$$

$$E\chi_{+1/2}Y = -f_s^{\frac{1+q}{2}} m\chi_{+1/2}Y + \left[ f_s^{1+q} \frac{1}{a^{1/2}}\left(\frac{\partial}{\partial r} + \frac{1}{r}\right) + f_s^{\frac{1+2q}{2}} \frac{1}{a^{1/2}} \frac{\kappa}{r} + \frac{1}{2}\frac{\partial}{\partial r}\left(f_s^{1+q} \frac{1}{a^{1/2}}\right) \right] \varphi_{+1/2}Y -$$
$$- f_s^{\frac{1+2q}{2}} \frac{1}{a^{1/2}} \frac{1}{r} \frac{\partial}{\partial \theta}(\varphi(r,\theta))_{-1/2}Y - \frac{1}{2} f_s^{\frac{1+2q}{2}} \frac{1}{r} \frac{\partial}{\partial \theta}\left(\frac{1}{a^{1/2}}\right) \varphi_{-1/2}Y + \frac{m_\varphi}{r\sin\theta} f_s^{\frac{1+2q}{2}} \left(1 - \frac{1}{a^{1/2}}\right) \varphi_{-1/2}Y. \quad (20)$$

Then in (17) - (20), we can get rid of the derivatives $\frac{\partial}{\partial\theta}(\chi(r,\theta))$ and $\frac{\partial}{\partial\theta}(\varphi(r,\theta))$.

To this end, the equation (17) is multiplied by $_{-1/2}Y(\theta)$, the equation (18) is multiplied by $_{+1/2}Y(\theta)$ and then they are summed. The same is done with the equations (19), (20). We obtain

$$E\varphi = f_s^{\frac{1+q}{2}} m\varphi - \left[ f_s^{1+q} \frac{1}{a^{1/2}}\left(\frac{\partial}{\partial r} + \frac{1}{r}\right) - f_s^{\frac{1+2q}{2}} \frac{1}{a^{1/2}} \frac{\kappa}{r} + \frac{1}{2}\frac{\partial}{\partial r}\left(f_s^{1+q} \frac{1}{a^{1/2}}\right) \right] \chi +$$
$$+ \frac{m_\varphi}{r\sin\theta} f_s^{\frac{1+2q}{2}} \left(1 - \frac{1}{a^{1/2}}\right) \frac{2\left(_{+1/2}Y\right)\left(_{-1/2}Y\right)}{\left(_{-1/2}Y\right)^2 + \left(_{+1/2}Y\right)^2} \chi. \quad (21)$$

---

[‡] As opposed to (13) here and below we removed the indexes $j, m_\varphi$ in notations $_{\mp 1/2}Y(\theta)$ for short.

$$E\chi = -f_s^{\frac{1+q}{2}} m\chi + \left[ f_s^{1+q} \frac{1}{a^{1/2}} \left( \frac{\partial}{\partial r} + \frac{1}{r} \right) + f_s^{\frac{1+2q}{2}} \frac{1}{a^{1/2}} \frac{\kappa}{r} + \frac{1}{2} \frac{\partial}{\partial r} \left( f_s^{1+q} \frac{1}{a^{1/2}} \right) \right] \varphi +$$

$$+ \frac{m_\varphi}{r \sin\theta} f_s^{\frac{1+2q}{2}} \left( 1 - \frac{1}{a^{1/2}} \right) \frac{2 \left( {}_{+1/2}Y \right) \left( {}_{-1/2}Y \right)}{\left( {}_{-1/2}Y \right)^2 + \left( {}_{+1/2}Y \right)^2} \varphi.$$

(22)

Equations (21), (22) can be used for a standard procedure to obtain effective potentials. The angle $\theta$ and the particle energy $E$ in this case are parameters.

Then, let us briefly recall the procedure of obtaining effective potentials. Below, the expressions will be written in the dimensionless variables $\rho = \frac{r}{l_c}$, $2\alpha = \frac{r_0}{l_c}$, $\varepsilon = \frac{E}{m}$, $l_c = \frac{\hbar}{mc}$ is the Compton wavelength of the Dirac particle.

From the equations (21), (22), we will obtain the second order equation for the function $\psi(\rho,\theta)$, proportional either to $\varphi(\rho,\theta)$ or to $\chi(\rho,\theta)$.

In the first case,

$$\psi(\rho,\theta) = \varphi(\rho,\theta) \exp\left( \frac{1}{2} \int A_1(\rho',\theta) d\rho' \right).$$

(23)

The equation for $\psi(\rho,\theta)$ has the form of the Schrödinger equation

$$\frac{\partial^2 \psi(\rho,\theta)}{\partial \rho^2} + 2\left( E_{Schr} - U_{eff}(\rho,\theta) \right) \psi(\rho,\theta) = 0.$$

(24)

In the equation (24),

$$E_{Schr} = \frac{1}{2} \left( \varepsilon^2 - 1 \right),$$

$$U_{eff}(\rho,\theta) = \frac{1}{4} \frac{\partial A_1(\rho,\theta)}{\partial \rho} + \frac{1}{8} A_1^2(\rho,\theta) - \frac{1}{2} B_1(\rho,\theta).$$

(25)

In the expressions (23), (25),

$$A_1(\rho,\theta) = -\frac{1}{B(\rho,\theta)} \frac{\partial B(\rho,\theta)}{\partial \rho} - A(\rho) - D(\rho),$$

$$B_1(\rho,\theta) = -B(\rho,\theta) \frac{\partial}{\partial \rho} \left( \frac{A(\rho,\theta)}{B(\rho,\theta)} \right) - C(\rho,\theta) B(\rho,\theta) + A(\rho,\theta) D(\rho,\theta).$$

(26)

In the expressions (26),

$$A(\rho,\theta) = -\frac{1}{f_s^{1+q}\frac{1}{a^{1/2}}}\left[f_s^{1+q}\frac{1}{a^{1/2}}\frac{1}{\rho} + f_s^{\frac{1+2q}{2}}\frac{1}{a^{1/2}}\frac{\kappa}{\rho} + \frac{1}{2}\frac{\partial}{\partial\rho}\left(f_s^{1+q}\frac{1}{a^{1/2}}\right)\right.$$

$$\left. + \frac{m_\varphi}{\rho\sin\theta}\left(1 - \frac{1}{a^{1/2}}\right)\frac{2\left(_{-1/2}Y\right)\left(_{+1/2}Y\right)}{\left(_{-1/2}Y\right)^2 + \left(_{+1/2}Y\right)^2}\right]. \tag{27}$$

$$B(\rho,\theta) = \frac{1}{f_s^{1+q}\frac{1}{a^{1/2}}}\left(\varepsilon + f_s^{\frac{1+q}{2}}\right). \tag{28}$$

$$C(\rho,\theta) = -\frac{1}{f_s^{1+q}\frac{1}{a^{1/2}}}\left(\varepsilon - f_s^{\frac{1+q}{2}}\right). \tag{29}$$

$$D(\rho,\theta) = -\frac{1}{f_s^{1+q}\frac{1}{a^{1/2}}}\left[f_s^{1+q}\frac{1}{a^{1/2}}\frac{1}{\rho} - f_s^{\frac{1+2q}{2}}\frac{1}{a^{1/2}}\frac{\kappa}{\rho} + \frac{1}{2}\frac{\partial}{\partial\rho}\left(f_s^{1+q}\frac{1}{a^{1/2}}\right)\right.$$

$$\left. - \frac{m_\varphi}{\rho\sin\theta}\left(1 - \frac{1}{a^{1/2}}\right)\frac{2\left(_{-1/2}Y\right)\left(_{+1/2}Y\right)}{\left(_{-1/2}Y\right)^2 + \left(_{+1/2}Y\right)^2}\right]. \tag{30}$$

The expression $F(\theta) = \dfrac{2\left(_{-1/2}Y\right)\left(_{+1/2}Y\right)}{\sin\theta\left[\left(_{-1/2}Y\right)^2 + \left(_{+1/2}Y\right)^2\right]}$ in (27), (30) can be represented as

$$F(\theta) = \frac{\left(P_l^{m_\varphi + \frac{1}{2}}(\cos\theta)\right)^2 - \left(\kappa - m_\varphi + \frac{1}{2}\right)^2\left(P_l^{m_\varphi - \frac{1}{2}}(\cos\theta)\right)^2}{\left(\kappa - m_\varphi + \frac{1}{2}\right)^2\left(P_l^{m_\varphi - \frac{1}{2}}(\cos\theta)\right)^2 + \left(P_l^{m_\varphi + \frac{1}{2}}(\cos\theta)\right)^2}$$

$$+ \frac{2\mathrm{ctg}\theta\left(\kappa - m_\varphi + \frac{1}{2}\right)P_l^{m_\varphi - \frac{1}{2}}P_l^{m_\varphi + \frac{1}{2}}(\cos\theta)}{\left(\kappa - m_\varphi + \frac{1}{2}\right)^2\left(P_l^{m_\varphi - \frac{1}{2}}(\cos\theta)\right)^2 + \left(P_l^{m_\varphi + \frac{1}{2}}(\cos\theta)\right)^2}. \tag{31}$$

Let us note again that the polar angle $\theta$ in the expressions (23) - (30) is the parameter varying in the interval $\theta \in [0, \pi]$. The particle energy $\varepsilon$ in the potential $U_{eff}$ (25) is a parameter as well.

The effective potentials $U_{eff}(\rho, \alpha, q, \kappa, l, m_\varphi, \varepsilon, \theta)$ determined by the expressions (25) - (30) have a cumbersome analytical form and are calculated in this paper by using the software package "Maple".

## 3. Analysis of the dependences $U_{eff}(\rho, \alpha, q, \kappa, l, m_\varphi, \varepsilon, \theta)$

The dependences $U_{eff}(\rho, \alpha, q, \kappa, l, m_\varphi, \varepsilon, \theta)$ were analyzed for the states of the Dirac particles enlisted in table 1.

Table 1. Dependences $F(\theta)$ for different states of the Dirac particle.

| № | | | | | $F(\theta)$ |
|---|---|---|---|---|---|
| 1 | $j = 1/2$ | $\kappa = -1$ | $l = 0$ | $m_\varphi = +1/2$ | -1 |
| 2 | $j = 1/2$ | $\kappa = -1$ | $l = 0$ | $m_\varphi = -1/2$ | +1 |
| 3 | $j = 1/2$ | $\kappa = +1$ | $l = 1$ | $m_\varphi = +1/2$ | +1 |
| 4 | $j = 1/2$ | $\kappa = +1$ | $l = 1$ | $m_\varphi = -1/2$ | -1 |
| 5 | $j = 3/2$ | $\kappa = -2$ | $l = 1$ | $m_\varphi = +3/2$ | -1 |
| 6 | $j = 3/2$ | $\kappa = -2$ | $l = 1$ | $m_\varphi = +1/2$ | $\dfrac{1-9\cos^2\theta}{1+3\cos^2\theta}$ |
| 7 | $j = 3/2$ | $\kappa = -2$ | $l = 1$ | $m_\varphi = -1/2$ | $-\dfrac{1-9\cos^2\theta}{1+3\cos^2\theta}$ |
| 8 | $j = 3/2$ | $\kappa = -2$ | $l = 1$ | $m_\varphi = -3/2$ | +1 |
| 9 | $j = 3/2$ | $\kappa = +2$ | $l = 2$ | $m_\varphi = +3/2$ | +1 |
| 10 | $j = 3/2$ | $\kappa = +2$ | $l = 2$ | $m_\varphi = +1/2$ | $-\dfrac{1-9\cos^2\theta}{1+3\cos^2\theta}$ |
| 11 | $j = 3/2$ | $\kappa = +2$ | $l = 2$ | $m_\varphi = -1/2$ | $\dfrac{1-9\cos^2\theta}{1+3\cos^2\theta}$ |
| 12 | $j = 3/2$ | $\kappa = +2$ | $l = 2$ | $m_\varphi = -3/2$ | -1 |

From the expressions (25) - (30) and from the data of table 1, it is seen that the effective potentials are symmetrical relative to the equatorial plane $(U_{eff}(\theta) = U_{eff}(\pi - \theta))$. It is also seen that the effective potentials have the only singularity near the radius $\rho = 2\alpha$.

According to (25) - (30), $m_\varphi$ and $F(\theta)$ are included into the effective potentials $U_{eff}(\rho, \theta)$ only in the form of the product $m_\varphi F(\theta)$. It follows therefrom and from table 1 that at the given value of the parameter $\kappa$ for $m_\varphi$, different only in signs, the potentials $U_{eff}(\rho, \theta)$ are the same.

For all the values $q$ as $\rho \to \infty$ $U_{eff}(\rho, \theta) \to 0$.

According to (25) - (30), the effective potentials have the following singularities near the radius of $\rho = 2\alpha$.

For $\theta \neq 0$, $\theta \neq \pi$

$$U_{eff}\big|_{\rho \to 2\alpha} = \frac{2^{2q-1}\alpha^{2q}m_\varphi^2}{(\rho-2\alpha)^{2(1+q)}} F^2(\theta). \tag{32}$$

For $\theta = 0$, $\theta = \pi$, the potential $U_{eff}\big|_{\rho \to 2\alpha}$ changes the sign and has the form of

$$U_{eff}(\sin\theta = 0)\big|_{\rho \to 2\alpha} = -\frac{2^{2q+1}\alpha^{2(1+q)}\varepsilon^2}{(\rho-2\alpha)^{2(1+q)}}. \tag{33}$$

Besides, there exists singularity

$$U_{eff}\big|_{\rho \to 2\alpha} = -\frac{N(q)}{(\rho-2\alpha)^2}, \tag{34}$$

where $N(q)$ is a numerical coefficient depending on $q$. In table 2, the values for $N$ are given for some negative $q$.

Table 2. The dependence $N(q)$ for negative q values.

| $q$ | -0.25 | -0.333 | -0.4 | -0.5 |
|---|---|---|---|---|
| $N$ | $\frac{1023}{8192} \approx \frac{1}{8.008}$ | $\frac{323}{2592} \approx \frac{1}{8.025}$ | $\frac{621}{5000} \approx \frac{1}{8.052}$ | $\frac{63}{512} \approx \frac{1}{8.127}$ |
| $q$ | -0.75 | -0.9 | -0.99 | -1 |
| $N$ | $\frac{943}{8192} \approx \frac{1}{8.687}$ | $\frac{33439}{320000} \approx \frac{1}{9.57}$ | $\frac{303940399}{3200000000} \approx \frac{1}{10.53}$ | $\frac{3}{32} \approx \frac{1}{10.667}$ |

It is seen from the table that in the entire allowed range of the negative q the coefficient $N < 1/8$. As $q \to 0$, the coefficient $N \to 1/8$.

The form of the expressions (32) - (34) shows that as $q > 0$ (oblate mass distribution) the leading are the expressions (32), (33). On the contrary, as $q < 0$ (prolate mass distribution) the leading is the expression (34). As $q \to 0$ the function $a(\rho,\theta) \to 1$ and the contribution (32) with $m_\varphi$ become vanishingly small. As $q \to 0$ both expressions (33), (34) as $q = 0$ become leading. In the limit $q = 0$, we obtain the expression for the leading term of the effective potential for the Schwarzschild metric in the vicinity of the event horizon [31]

$$U_{eff}^S\big|_{\rho \to 2\alpha} = -\frac{1/8 + 2\alpha^2\varepsilon^2}{(\rho-2\alpha)^2}. \tag{35}$$

In (35), the numerator is $> 1/8$. In this case the conditions of a quantum-mechanic "fall" of the Dirac particles to the event horizon $\rho = 2\alpha$ are implemented [21], [31].

### 3.1 Oblate mass distribution $q \in (0, \infty)$

For $\theta \neq 0$, $\theta \neq \pi$, the leading term $U_{eff}$ (25) has the form (32)

$$U_{eff}\big|_{\rho \to 2\alpha} = \frac{2^{2q-1} \alpha^{2q} m_\varphi^2}{(\rho - 2\alpha)^{2(1+q)}} F^2(\theta). \tag{36}$$

For $\theta = 0$, $\theta = \pi$, the leading term $U_{eff}$ changes the sign and has the form (33)

$$U_{eff}(\sin\theta = 0)\big|_{\rho \to 2\alpha} = -\frac{2^{1+2q} \alpha^{2(1+q)} \varepsilon^2}{(\rho - 2\alpha)^{2(1+q)}}. \tag{37}$$

If $F(\theta) \neq 0$, the expression (36) represents an infinitely large repulsive barrier. The barrier value sharply increases with the growth of $q$. On the contrary, as $\theta = 0$, $\theta = \pi$ the leading term $U_{eff}\big|_{\rho \to 2\alpha}$ represents a depth-unlimited potential well. At some values of $\kappa, l, m_\varphi$ and at certain values of $\theta_i$ the function $F(\theta)$ and the expression (36) become equal to zero. In table 1 $F(\theta_i) = 0$ as

$j = 3/2$, $\kappa = -2$, $m_\varphi = \pm 1/2$, $\cos\theta_i = \pm 1/3$; $j = 3/2$, $\kappa = +2$, $m_\varphi = \pm 1/2$, $\cos\theta_i = \pm 1/3$.

In this case as $\theta = \theta_i$ and $q > q^* \approx \sqrt{2}$ the leading term $U_{eff}(\theta = \theta_i)\big|_{\rho \to 2\alpha}$, as before, represents an infinitely high repulsive barrier of the following form

$$U_{eff}(\theta = \theta_i)\big|_{\rho \to 2\alpha} = \frac{L}{(\rho - 2\alpha)^2}. \tag{38}$$

In (38), $L$ is a coefficient increasing with the growth of $q$. As $q \approx q^*$, the repulsive barrier disappears $(L = 0)$. In the interval $0 < q < q^*$, there appears a potential well of the following form

$$U_{eff}(\theta = \theta_i)\big|_{\rho \to 2\alpha} = -\frac{L_1}{(\rho - 2\alpha)^2} \tag{39}$$

with the coefficient $L_1 < 1/8$.

Fig. 1 represents one-dimensional (with fixed values of the angle $\theta$) and two-dimensional dependences $U_{eff}(\rho, \theta)$ for some states of a Dirac particle with positive values $m_\varphi$ (Table 1, items 1, 3, 5, 6, 9, 10). The dependences $U_{eff}(\rho, \theta)$ for the states with the appropriate negative values $m_\varphi$ do not quantitatively vary.

Let us discuss the form of $U_{eff}(\rho, \theta)$ in the interval $\theta \neq 0$, $\theta \neq \pi$. For state 1 with $l = 0$, the effective potential near the outer naked singularity $\rho = 2\alpha$ represents an infinitely large repulsive barrier with a subsequent potential well. The finite depth of the well increases while moving from the equator $\left(\theta = \frac{\pi}{2}\right)$ towards the poles $(\theta = 0, \theta = \pi)$.

For the states with $l \neq 0$ and $F(\theta) \neq 0$, the depth of the potential well is essentially greater at positive $\kappa$. For certain states, the potential well disappears in the equatorial zone but with the motion towards the poles it appears at certain values of $\theta = \theta_{min}$ and $\theta = \pi - \theta_{min}$. The finite depth of the well increases while moving along the angle $\theta$ towards the poles.

For states 6, 10 with $l \neq 0$ and $F(\theta_i) = 0$ as $\theta = \theta_i$ and as $0 < q < q^*$ the positive repulsive barrier disappears. Instead, an infinitely deep potential well (39) appears.

For all the examined cases at the poles $(\theta = 0, \theta = \pi)$ there exists an infinitely deep potential well (37). Variations in $q$ (fig. 2) and $\varepsilon$ (fig. 3, 4) do not qualitatively change the nature of the dependences $U_{eff}(\rho, \theta)$.

### 3.2 Prolate mass distribution $q \in (-1, 0)$

In this case, the leading term $U_{eff}\big|_{\rho \to 2\alpha}$ has the form (34)

$$U_{eff}\big|_{\rho \to 2\alpha} = -\frac{N(q)}{(\rho - 2\alpha)^2}, \qquad (40)$$

Earlier, we have determined (see table 2) that for negative $q$ the coefficient $N < 1/8$. At that, in the limit $q \to 0$, the coefficient $N \to 1/8$. However, at modulo small negative $q$'s, an essential contribution of the expression (33) is added to the expression (34). For $q < 0$ when $q = q_{lim}$ ($|q_{lim}| \ll 1$), the function $U_{eff}\big|_{\rho \to 2\alpha}$ approaches $-\frac{1}{8}\frac{1}{(\rho - 2\alpha)^2}$. With the following decrease in $|q|$, the potential $U_{eff}$ tends to Schwarzschild limit (35). The value $q_{lim}$ in accordance with (33), (34) depends on the parameters of the q-metric (1) and the particle energy $\varepsilon$. The numerical values $q_{lim}$ can be determined by means of exact quantum-mechanical calculations only.

In the interval $q_{lim} < q < 0$, the condition of a quantum-mechanic "fall" of the Dirac particle to the outer singularity $\rho = 2\alpha$ is implemented.

According to quantum mechanics, (see, for instance, [21]), the attractive singular potential (40) for a prolate mass distribution as $-1 < q < q_{lim}$ enables the possibility of existence of stationary bound states of quantum-mechanical particles.

Fig. 5 presents two-dimensional dependences $U_{eff}(\rho,\theta)$ as $q=-0.5$ for the states of the Dirac particle considered in fig.1.

For the states with $l \neq 0$ in front of the potential well (40) there exist essential repulsive barriers. In all cases, the heights of the barriers notably depend on a polar angle $\theta$.

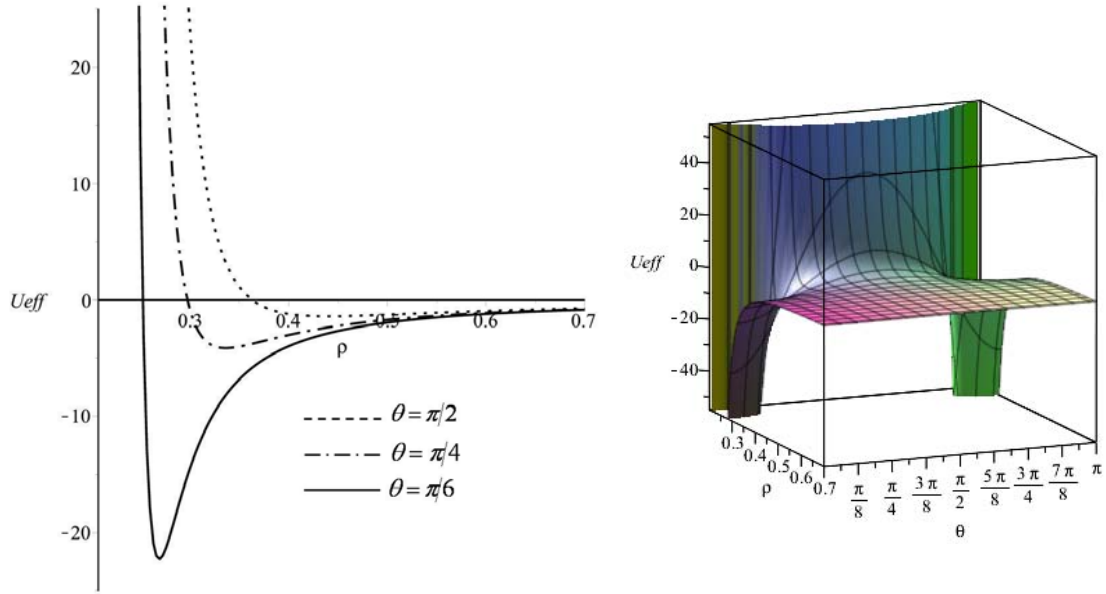

State 1: $\kappa=-1,\ l=0,\ m_\varphi=1/2$

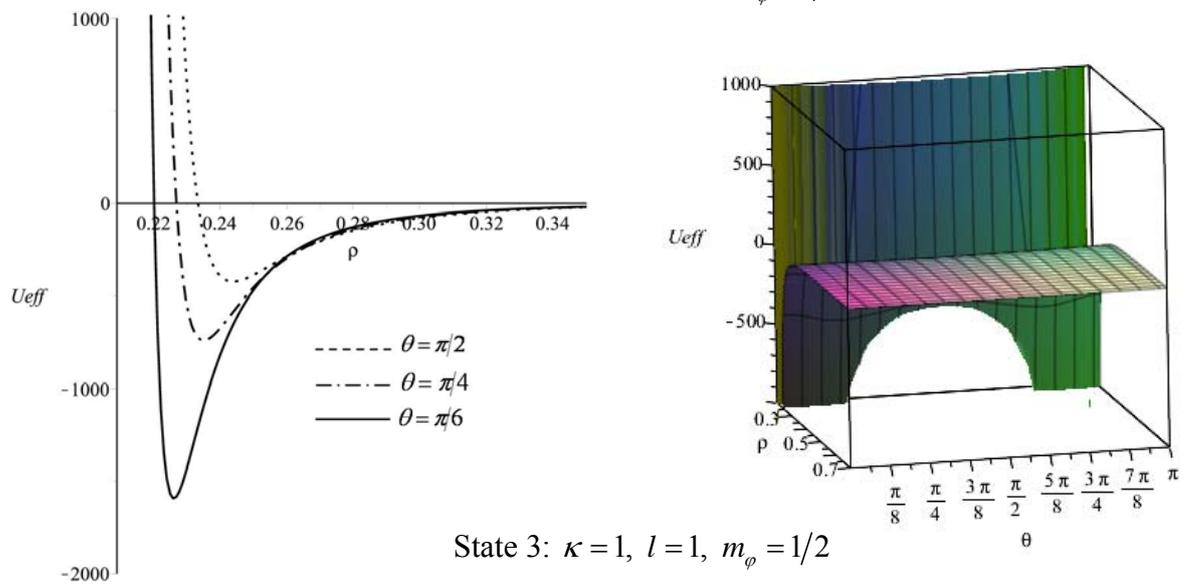

State 3: $\kappa=1,\ l=1,\ m_\varphi=1/2$

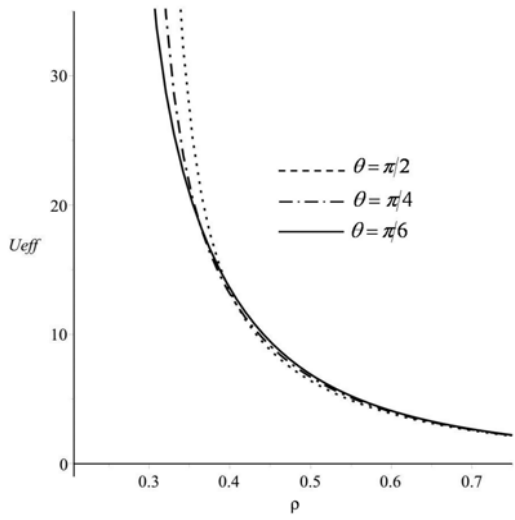
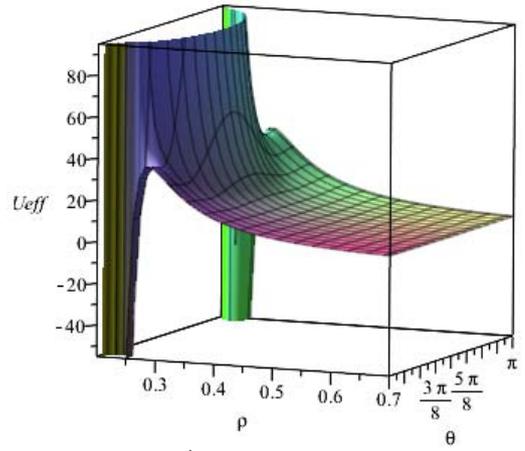

State 5: $\kappa = -2$, $l = 1$, $m_\varphi = 3/2$

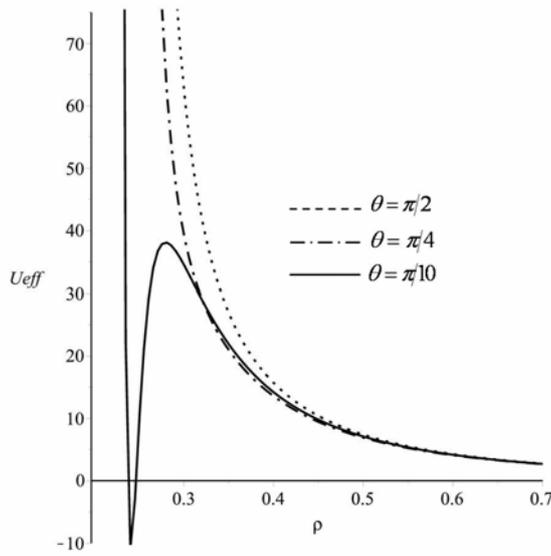
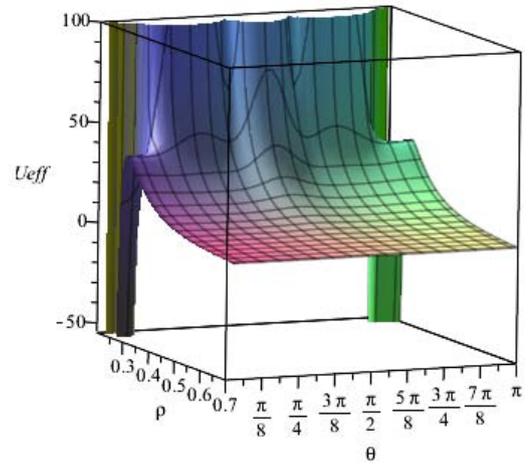

State 6: $\kappa = -2$, $l = 1$, $m_\varphi = 1/2$

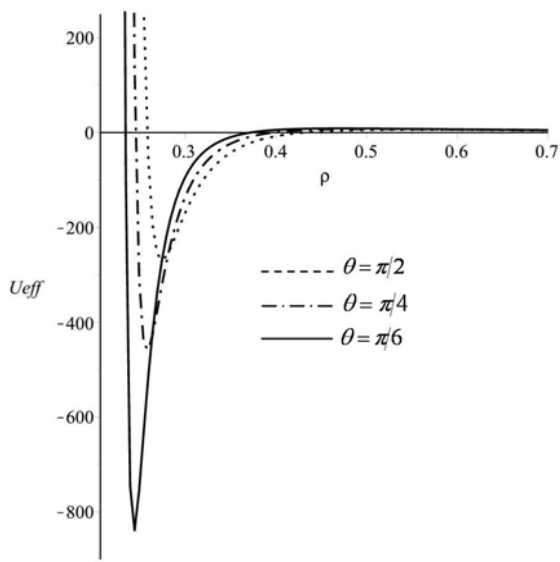
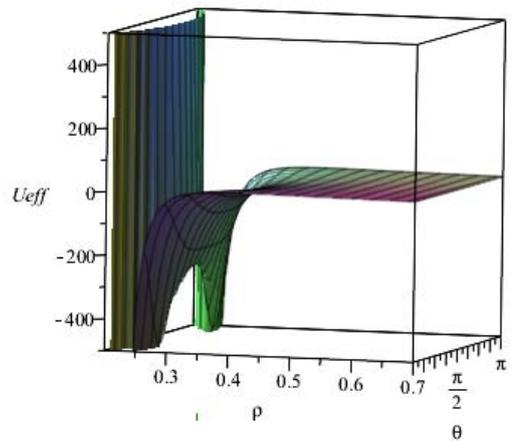

State 9: $\kappa = 2$, $l = 2$, $m_\varphi = 3/2$

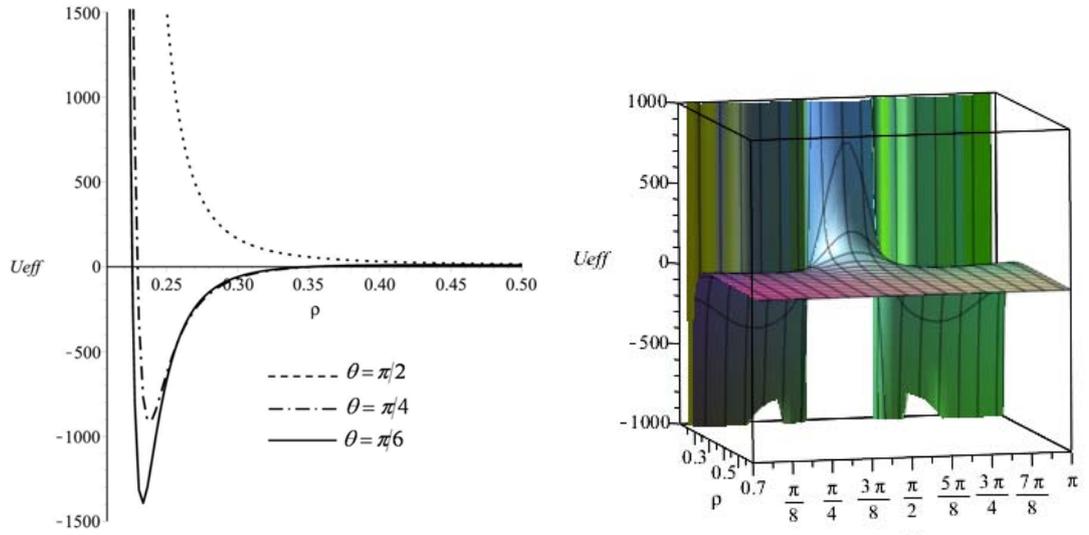

State 10: $\kappa = 2,\ l = 2,\ m_\varphi = 1/2$

Fig.1. One-dimensional and two-dimensional dependences as $\alpha = 0.1,\ q = 1,\ \varepsilon = 1$

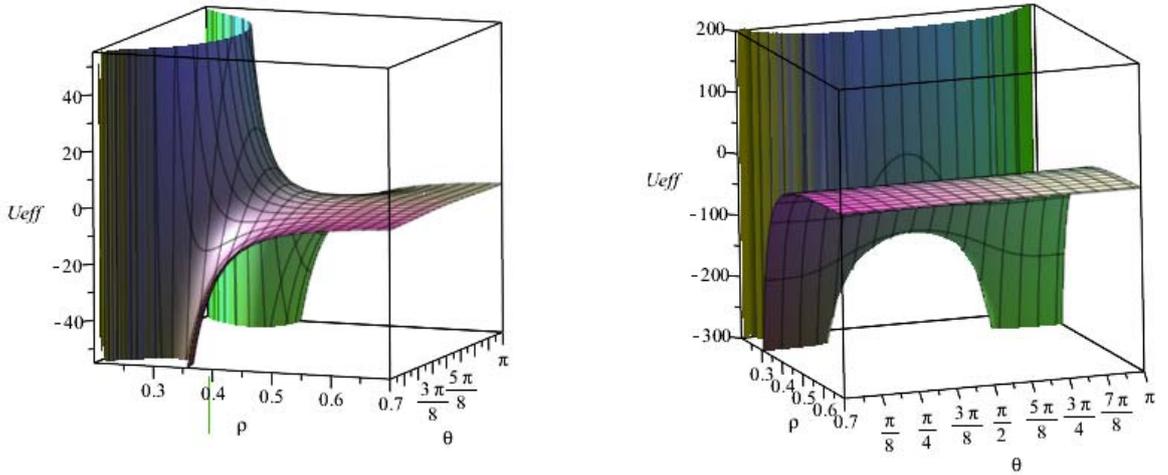

State 1: $\kappa = -1,\ l = 0,\ m_\varphi = 1/2$      State 3: $\kappa = 1,\ l = 1,\ m_\varphi = 1/2$

Fig. 2. Two-dimensional dependences $U_{eff}(\rho,\theta)$ as $\alpha = 0.1,\ q = 2,\ \varepsilon = 1$

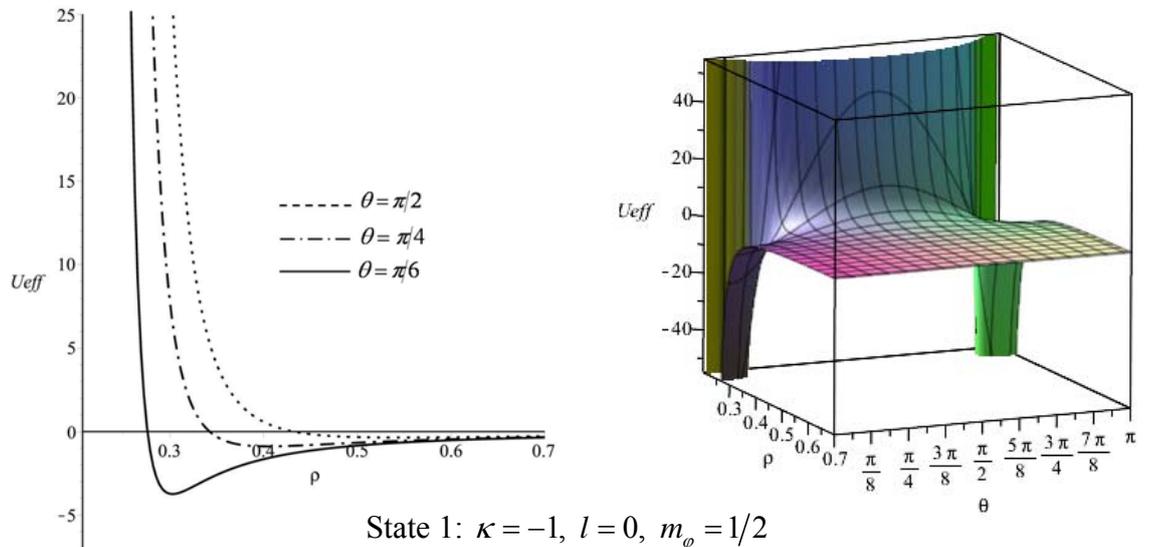

State 1: $\kappa = -1,\ l = 0,\ m_\varphi = 1/2$

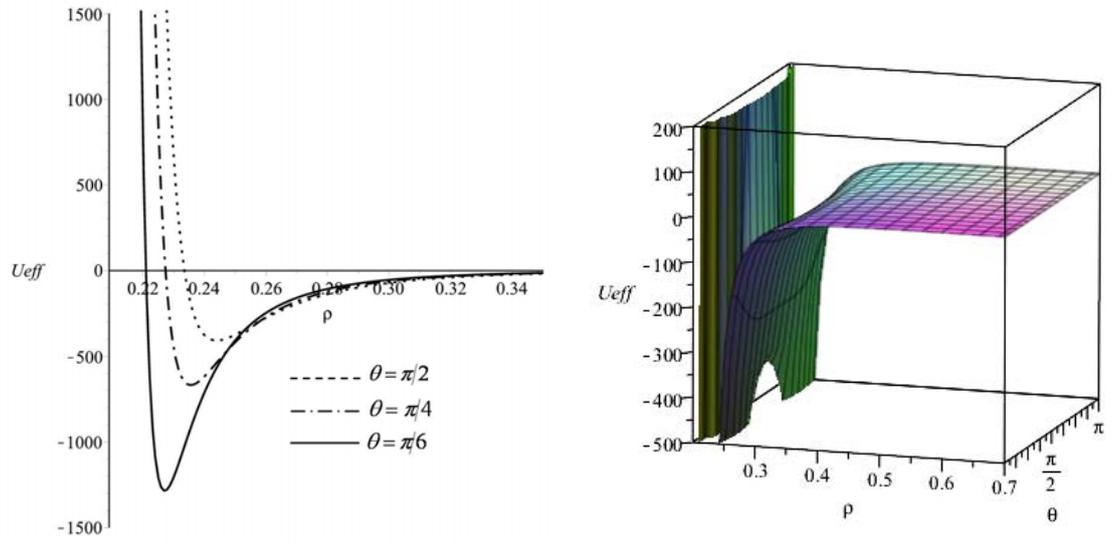

State 3: $\kappa = 1, \ l = 1, \ m_\varphi = 1/2$

Fig.3. One-dimensional and two-dimensional dependences $U_{eff}(\rho, \theta)$ as

$\alpha = 0.1, \ q = 1, \ \varepsilon = 0.8$

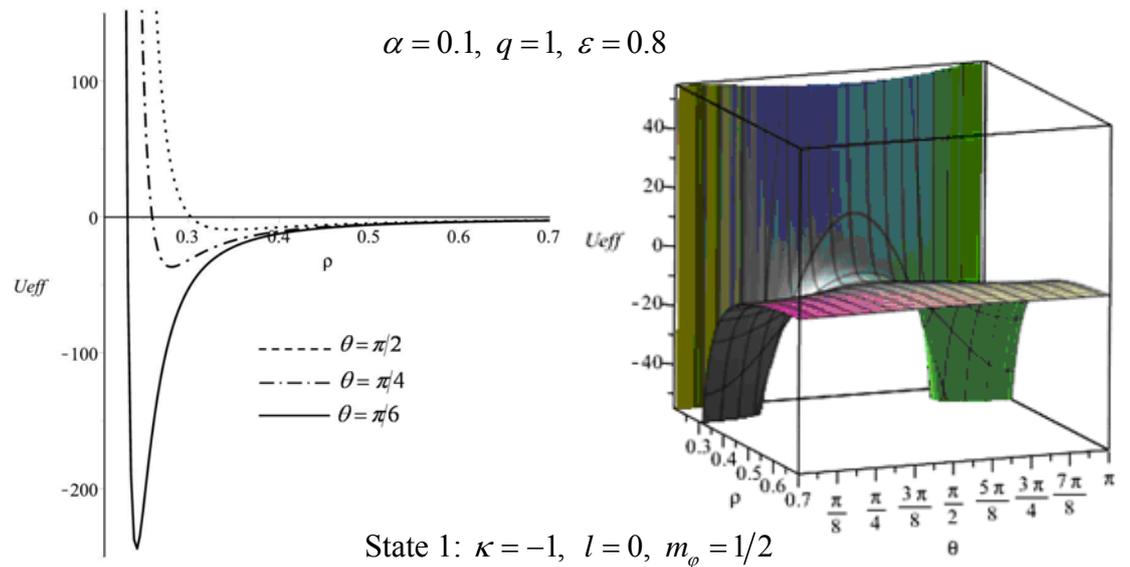

State 1: $\kappa = -1, \ l = 0, \ m_\varphi = 1/2$

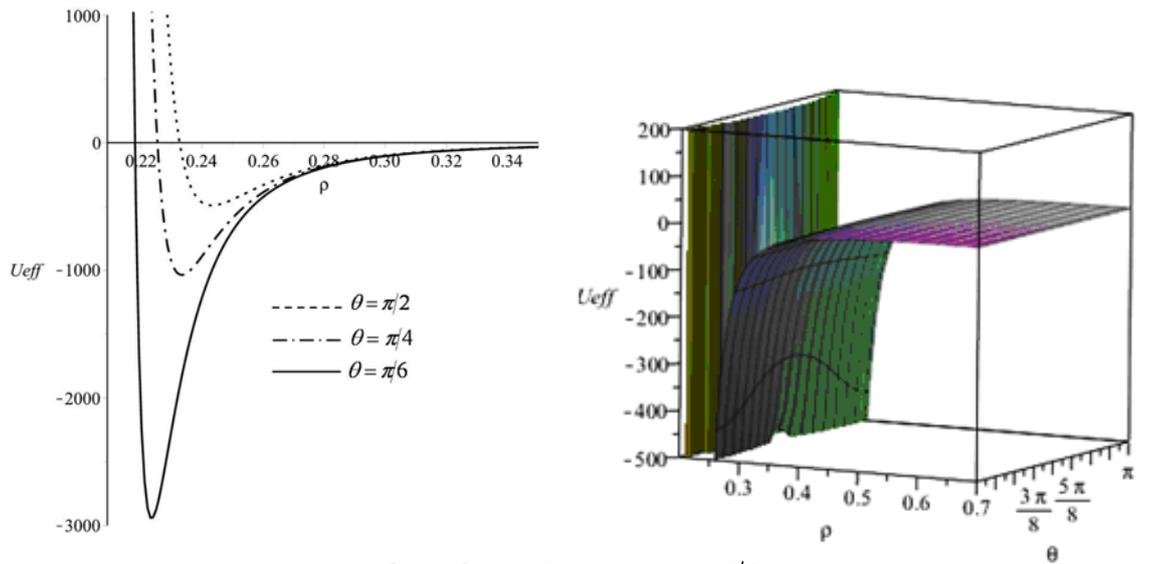

State 3: $\kappa = 1, \ l = 1, \ m_\varphi = 1/2$

Fig. 4. One-dimensional and two-dimensional dependences $U_{eff}(\rho, \theta)$ as

$\alpha = 0.1, \ q = 1, \ \varepsilon = 1.5$

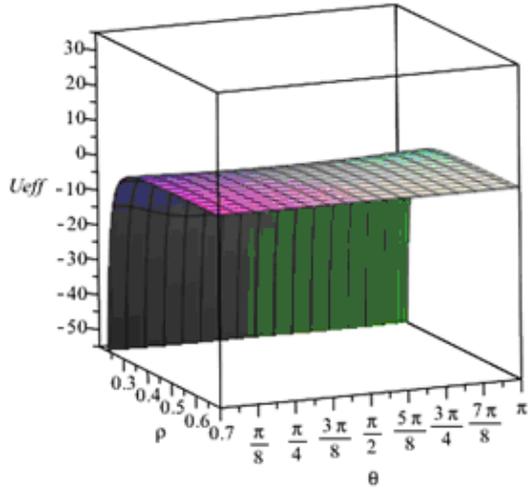

State 1: $\kappa = -1,\ l = 0,\ m_\varphi = 1/2$

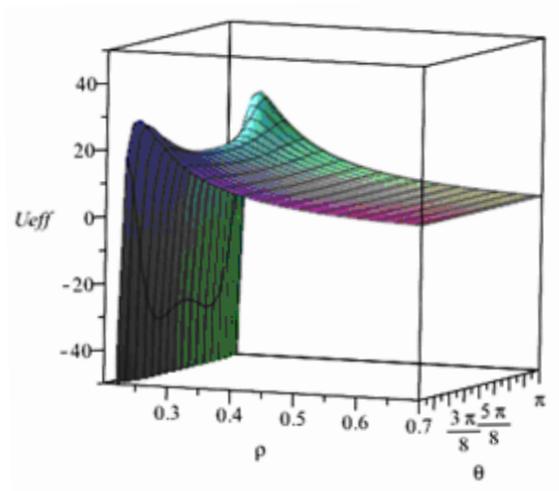

State 3: $\kappa = 1,\ l = 1,\ m_\varphi = 1/2$

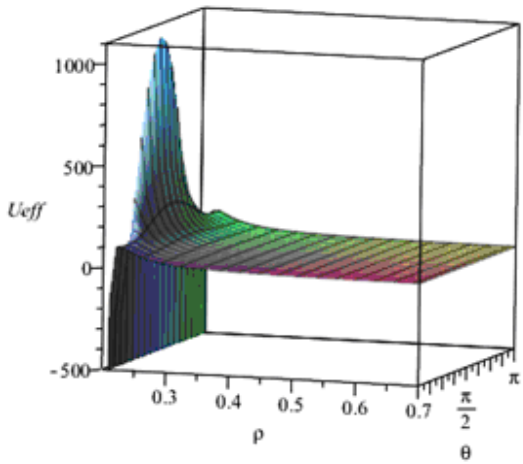

State 5: $\kappa = -2,\ l = 1,\ m_\varphi = 3/2$

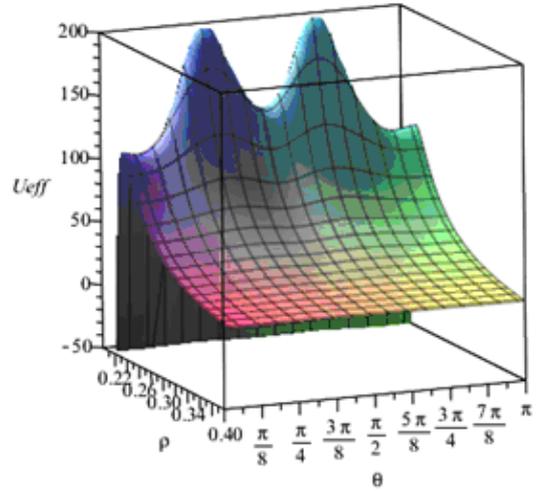

State 6: $\kappa = -2,\ l = 1,\ m_\varphi = 1/2$

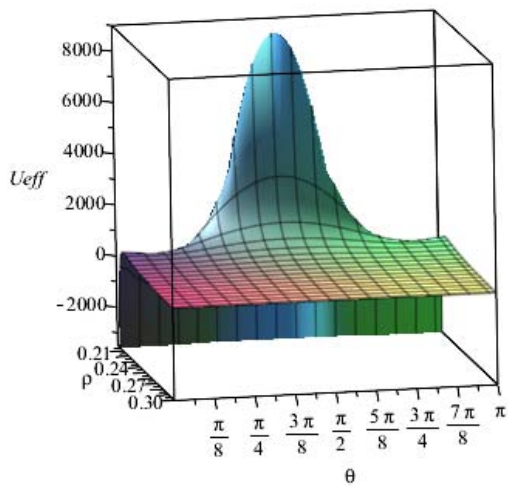

State 9: $\kappa = 2,\ l = 2,\ m_\varphi = 3/2$

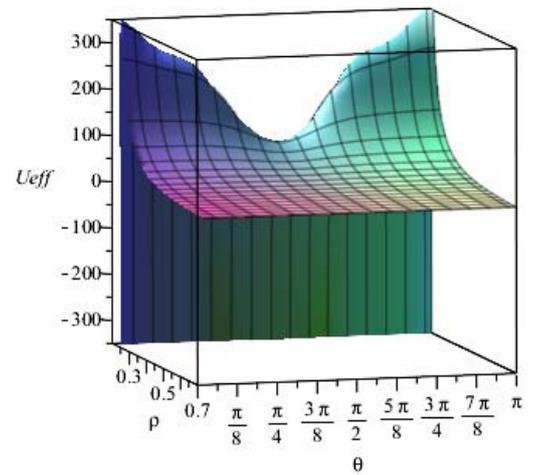

State 10: $\kappa = 2,\ l = 2,\ m_\varphi = 1/2$

Fig. 5. Two-dimensional dependences $U_{eff}(\rho,\theta)$ as $\alpha = 0.1,\ q = -1/2,\ \varepsilon = 1$

**4. Cosmic censorship and a $q$ - metric**

The hypothesis of cosmic censorship proposed more than 40 years ago [22] prohibits the existence of singularities, not shielded by event horizons. However, there is still no complete proof of this hypothesis. Many researchers examine, along with black holes, the formation of naked singularities, their stability and distinctive features during experimental observations (see, for instance, [23] - [28]). It is shown in [29] that there exist static metrics with time-like singularities which prove out to be completely non-singular when examining quantum mechanics of spinless particles. In [30], these results are validated as applied to the motion of half-spin quantum-mechanical particles in the field of the Reissner-Nordström naked singularity. For any Dirac particle irrespective of availability and sign of its electric charge, the Reissner-Nordström naked singularity is separated by an infinitely large repulsive barrier.

$$U_{eff}\big|_{\rho \to 0} = \frac{3}{8\rho^2} + O\left(\frac{1}{\rho}\right). \tag{41}$$

According to the vivid expression of the authors of [29], the existence of a repulsive barrier covering the singularity does not threat the cosmic censorship.

Let us discuss the naked singularities of the static q-metric in the similar way.

In case of the oblate mass distribution $\left(q \in (0,\infty)\right)$, we see from (36), (38) that the naked singularities of a $q$-metric are covered with infinitely large repulsive barriers which agrees with the hypothesis of the cosmic censorship. The exception are the poles and, as $0 < q < q^*$, some points $\theta_i$ for the states of a particle with $j \geq 3/2$. In this case, in points $\theta = \theta_i$ instead of a repulsive barrier there exists a potential well of the form (39) with the coefficient $L_1$, allowing the existence of stationary bound states of Dirac particles. At the poles, there exists an infinitely deep potential well (37). The effect of the poles and the points $\theta = \theta_i$ as $j \geq 3/2$ and $0 < q < q^*$ on the conclusion of compliance with the hypothesis of cosmic censorship must be determined in more accurate quantum-mechanical calculations of solution to the Dirac equation.

In case of the prolate mass distribution $\left(q \in (-1,0)\right)$ along the axial axis, the naked singularities for some intervals $\theta$ are not covered with a repulsive barrier. However, the view of the leading term (40) as $-1 < q < q_{\lim}$ testifies to the existence possibility of stationary bound states of a Dirac particle.

In the interval $q_{\lim} \leq q < 0$, there are implemented the conditions of a quantum-mechanical "fall" of a Dirac particle to the outer singularity $\rho = 2\alpha$.

**Conclusions**

The quantum-mechanical motion of half-spin particles was examined in the field of naked singularities of the static q-metric formed by mass distribution with a quadrupole moment. The analysis was performed by means of the method of effective potentials of the Dirac equation generalized for the case when radial and angular variables are not separated. In order to obtain effective potentials, the self-conjugate Dirac Hamiltonian in the field of the naked singularities of the q-metric was determined.

The q-metric is transformed to the Schwarzschild metric as $q = 0$. The leading term of the effective potential for the Schwarzschild metric near the event horizon is

$$U_{eff}^{S}\Big|_{\rho \to 2\alpha} = -\frac{\frac{1}{8}+2\alpha^{2}\varepsilon^{2}}{(\rho-2\alpha)^{2}}. \tag{42}$$

The expression (42) testifies to the fact that the motion of the Dirac particle in the Schwarzschild field is implemented in the mode of a "fall" to the event horizon [31].

The transition from the spherical-symmetric to the deformed mass distribution in the source of the gravitational field leads to essential variations in the motion of the Dirac particle. The motion conditions of a half-spin particle strongly differ as well subject to the form of mass distribution.

For the prolate mass distribution along the axial axis of mass distribution $(-1 < q < q_{\lim},\ |q_{\lim}| \ll 1)$, the view of the leading term of the effective potential near the outer naked singularity (40) with the coefficient $N$ in table 2 testifies to the existence possibility of stationary bound states of half-spin particles. An analogy can be the existence of the energy spectrum of electrons in the field of the singular Coulomb potential in hydrogen-like atoms with $Z < 137$ (the Sommerfeld formula). As $q_{\lim} \leq q < 0$, there exist the conditions for a quantum-mechanical "fall" of particles to the outer singularity $\rho = 2\alpha$.

For the oblate mass distribution along the axial axis $(0 < q < \infty)$, the effective potential near the outer naked singularity has a more complicate form. The naked singularities are separated by infinitely high repulsive barriers (36), (38) with the subsequent transition to a potential well whose depth increases while moving along the angle from the equator to the poles. For the state with $l \neq 0$, the potential well can appear at a certain value of $\theta_{\min}$ and $\pi - \theta_{\min}$. Generally speaking, such a potential agrees with the cosmic censorship since the naked singularities of the q-metric are shielded by an impenetrable quantum-mechanical barrier. The exception are the poles and, as $0 < q < q^*$, some points $\theta_i$ for the states of the particle with $j \geq 3/2$. In these points, the leading terms of the effective potentials are attractive potentials (37), (39). The potentials with the leading term (39) allow the existence possibility of stationary bound states of Dirac particles. Only at the poles $\theta = 0$, $\theta = \pi$ there exist conditions for a quantum-mechanical "fall" of the particle to the outer singularity $\rho = 2\alpha$. The effect of the poles and the finite number of points $\theta = \theta_i$ on the penetrability characteristics of the barriers must be evaluated in more accurate quantum-mechanical calculations of the solution to the Dirac equation in the field of the naked singularities of the q-metric. But we can already say that these characteristics will be changed insignificantly.

Thereby for the oblate mass distribution $(0 < q < \infty)$, the naked singularities of the q-metric are separated from the Dirac particle by infinitely large repulsive barriers that is conformed with cosmic censorship. For a prolate mass distribution $(-1 < q < q_{\lim}, |q_{\lim}| \ll 1)$, the singularity of the effective potentials enables the possibility of existence of stationary bound states of half-spin particles. The conditions of a quantum-mechanic "fall" of the particles to the outer singularity $\rho = 2\alpha$ are implemented as $q_{\lim} < q < 0$ only.

**Acknowledgements**

The authors would like to thank A.L. Novoselova for the essential technical support while elaborating the paper.